\documentstyle[preprint,eqsecnum,cite,aps]{revtex}
\begin{document}
\title{Relativistic Corrections to the Aharonov-Bohm Scattering}
\author{M. Gomes, J. M. C. Malbouisson\footnote{On leave of absence
from Instituto de F\'\i sica, Universidade Federal da Bahia, Salvador,
40210-340, Brazil.} and A. J. da Silva} \address{Instituto de F\'\i
sica, Universidade de S\~ao Paulo, Caixa Postal 66318,\\ 05315--970,
S\~ao Paulo, SP, Brazil.}

\maketitle
\begin{abstract}
We determine the $|{\bf p}|/m$ expansion of the two body scattering
amplitude of the quantum theory of a Chern-Simons field minimally coupled to
a scalar field with quartic self-interaction. It is shown that the existence
of a critical value of the self-interaction parameter for which the
2-particle amplitude reduces to the Aharonov-Bohm one is restricted to the
leading, nonrelativistic, order. The subdominant terms correspond to
relativistic corrections to the Aharonov-Bohm scattering.
\end{abstract}

\noindent
{PACS: 03.65.Bz,03.70.+k,11.10.Kk}

\narrowtext

\section{INTRODUCTION}

The Aharonov-Bohm (AB) effect [$\citen{ab}$], the scattering of
charged particles by a flux line, is one of the most fascinating problems of
planar nonrelativistic (NR) quantum dynamics which has recently regained
great interest due to its connection to the physics of anyons [$\citen{lm}$-$
\citen{aro}$]. It has exact solution but if the scattering potential is
treated as a perturbation, due to its singular nature, divergences occur in
the perturbative series and so an appropriate renormalization is necessary
to get the correct expansion. Moreover, through perturbative
renormalization, the AB potential becomes related to the delta function,
contact, interaction. These issues have been discussed from the quantum
mechanical (first-quantized) viewpoint [$\citen{qm}$] and also by using a
field theoretical (second-quantized) approach.

In the latter case, one considers a NR scalar field minimally coupled to a
Chern-Simons (CS) gauge field [$\citen{hag2}$] and includes a quartic
self-interaction [$\citen{jac}$] in such a manner that the Galilean
invariant Lagrangian density [$\citen{bl}$] is given by

\begin{equation}
{\cal L}_{NR}=\psi ^{*}\left( iD_t+\frac{{\bf D}^2}{2m}\right) \psi -\frac{%
v_0}4(\psi ^{*}\psi )^2+\frac \Theta 2\partial _t{\bf A}\times {\bf A}%
-\Theta A_0{\bf \nabla }\times {\bf A}\;.  \label{lagranNR}
\end{equation}
This Lagrangian deeply differs from a relativistic one in the particle
kinetics which is manifest in the NR free particle propagator $\Delta
_{NR}(\omega ,{\bf k})=i\left[ \omega -{\bf k}^2/2m+i\varepsilon \right]
^{-1}$.Up to one loop, the 2-particle scattering amplitude,
calculated in the center of mass (CM) frame, is

\begin{equation}
{\cal A}_{NR}^{(1)}=-v_0-i\frac{2e^2}{m\Theta }\cot \theta +\frac m{8\pi }\left( v_0^2-\frac{4e^4}{m^2\Theta ^2}%
\right) \left[ \ln \left( \frac{\Lambda _{NR}^2}{{\bf p}^2}\right) +i\pi
\right] \;,  \label{A1NRcof}
\end{equation}
where $\Lambda _{NR}$ is a nonrelativistic ultraviolet cutoff . The
renormalization is implemented by redefining the self-coupling constant, $%
v_0=v+\delta v$, so that the total renormalized nonrelativistic amplitude is given, up to order $e^4$, by

\begin{equation}
{\cal A}_{NR}=-v-i\frac{2e^2}{m\Theta }\cot \theta +\frac m{8\pi }\left( v^2-%
\frac{4e^4}{m^2\Theta ^2}\right) \left[ \ln \left( \frac{\mu ^2}{{\bf p}^2}%
\right) +i\pi \right] \;,  \label{ANRren}
\end{equation}
where $\mu $ is an arbitrary mass scale, introduced by the renormalization,
that breaks the scale invariance of the amplitude [$\citen{bl}$].

At the critical values $v_c^{\pm }=\pm 2e^2/m|\Theta |$, the one loop
scattering vanishes, scale invariance is restored and by choosing the $%
v_c^{+}$ value, corresponding to a repulsive contact interaction, the
amplitude reduces, after multiplying by the appropriate kinematical factor,
to the Aharonov-Bohm amplitude for identical particles which is given by [$%
\citen{ab},\citen{bl}$] 
\begin{equation}
{\cal F}_{{\rm AB}}(|{\bf p}|,\theta )=-i\sqrt{\frac \pi {|{\bf p}|}}%
\,\alpha \,\left[ \cot \theta -i\,{\rm sgn\,}(\alpha )\right] +{\cal O}%
\left( \alpha ^3\right) \;,  \label{AB}
\end{equation}
where $\alpha =e^2/2\pi \Theta $. The existence of scale invariance at the
critical self-coupling has been explicitly verified up to three loop level
using differential regularization [$\citen{flr}$] and was recently proven to
hold in all orders of perturbation theory [$\citen{kim}$]. Also, starting
from the relativistic action for scalar self-interacting field minimally
coupled to a CS gauge field, it was shown that the nonrelativistic limit
of the one loop renormalized 2-particle scattering reduces
to the nonrelativistic scale invariant one for the same critical values [$%
\citen{boz}$].

This letter is concerned with the question to what extent the relativistic corrections preserve such criticality. Using an intermediate cutoff procedure [$%
\citen{go}$], which allows the determination of the $|{\bf p}|/m$ expansion
of the quantum amplitudes, we calculate the 1-loop particle-particle CM
scattering amplitude, for low external momenta, up to order ${\bf p}^2/m^2$.
The leading term of the $|{\bf p}|/m$ expansion coincides with the result of
Ref. [$\citen{boz}$] whereas the subdominant parts do not vanish at the
critical self-interaction values and so represent relativistic corrections
to the Aharonov-Bohm scattering.

\section{The Relativistic Model}

We consider a charged self-interacting scalar field in 2 + 1 dimensions
minimally coupled to a Chern-Simons gauge field described by the Lagrangian
density

\begin{equation}
{\cal L}=(D_\mu \phi )^{*}(D^\mu \phi )-m^2\phi ^{*}\phi -\frac \lambda 4
(\phi ^{*}\phi )^2+\frac \Theta 2\epsilon _{\sigma \mu \nu }A^\sigma
\partial ^\mu A^\nu -\frac \xi 2(\partial _iA^i)^2\;,  \label{lagran}
\end{equation}
where $D_\mu =\partial _\mu -ieA_\mu $ is the covariant derivative, $
\epsilon _{\sigma \mu \nu }$ is the fully antisymmetric tensor normalized to 
$\epsilon _{012}=+1$, the Minkowski metric signature is (1, -1, -1), the
units are such that $\hbar =c=1$ and repeated greek indices sum from 0 to 2 while repeated latin indices sum from 1 to 2. The choice of the Coulomb gauge fixing,
the same used in Ref. [$\citen{bl}$], is dictated by convenience in
discussing the nonrelativistic limit. It generates, in the Landau limit $
(\xi \rightarrow \infty )$, a virtual gauge field propagator which is
independent of $k^0$ and totally antisymmetric in the Minkowski indices with
the only nonvanishing components given by

\begin{equation}
D_{0i}(k)=-D_{i0}(k)=\frac 1\Theta \frac{\epsilon _{ij}k^j}{{\bf k}^2}\;,
\label{propGFL}
\end{equation}
where $\epsilon _{ij}=\epsilon _{0ij}$. The free propagator of the bosonic
matter field is the usual Feynman propagator $\Delta (p)=i\left[
p^2-m^2+i\varepsilon \right] ^{-1}$ and the vertex factors are $-i\lambda $
for the self-interaction vertex and $-ie(p+p^{\prime })^\mu $ and $%
2ie^2g^{\mu \nu }$ for the trilinear and the seagull vertices that always
arise from minimal coupling with a scalar field. With $V$ and $N$ denoting
the numbers of vertices and external lines, the degree of superficial
divergence of a generic graph is given by $d(G)=3-\frac 12N_\phi
-N_A-V_{\phi ^4}$ and, thus, we must expect to face divergences in calculating
radiative corrections. In the sequel we shall analyse these corrections at
one loop level.

Owing to charge conjugation and the antisymmetry of the gauge field
propagator (\ref{propGFL}), the only non-vanishing contribution to the
1-loop particle self-energy is an infinite constant coming from the
tadpole graph which can be absorbed into the definition of the
physical mass $m$.

The vacuum polarization shown in Fig. 1 is, naturally,
independent of the gauge fixing chosen and dimensional renormalization gives

\begin{equation}
\Pi _{\mu \nu }^{(1)}(q)=-\frac{ie^2}{2\pi }m\left( 1-\int_0^1dx\sqrt{%
1-q^2x(1-x)/m^2}\right) \frac 1{q^2}\left[ q^2g_{\mu \nu }-q_\mu q_\nu
\right] \;,  \label{vacpol}
\end{equation}
which reduces, in the low momentum regime $\left| q^2\right| \ll m^2$ , to

\begin{equation}
\Pi _{\mu \nu }^{(1)}(q)\simeq -\frac{ie^2}{24\pi m}\left( 1+\frac{q^2}{20m^2%
}\right) \left[ q^2g_{\mu \nu }-q_\mu q_\nu \right] \;.  \label{vacpolLQ}
\end{equation}
It should be noticed that using a cutoff regulator one gets explicitly the
linearly divergent part but the finite one is the same as above.

The 1-loop correction to the trilinear gauge coupling vertex, shown in Fig.
2, is finite. (The other possible 1-loop contribution, which has one
trilinear and one self-interaction vertex inserted in a particle loop, is
null by charge conjugation.) The sum of the two first parcels, the
contribution involving the seagull vertex, is exactly given, for external
particles legs in the mass shell, by

\begin{equation}
\Gamma _{3(S)}^{(1)0}=0\;\;{\rm and}\;\;\Gamma _{3(S)}^{(1)l}=\frac{-e^3}{%
2\pi \Theta }\epsilon _{ij}g^{jl}\left[ \frac{p^i\sqrt{m^2+{\bf p}^2}}{\sqrt{%
m^2}+\sqrt{m^2+{\bf p}^2}}-\frac{p^{\prime \,i}\sqrt{m^2+{\bf p}^{\prime \,2}%
}}{\sqrt{m^2}+\sqrt{m^2+{\bf p}^{\prime \,2}}}\right] \;.  \label{gama3S}
\end{equation}

\noindent On the other hand, the triangle graph can be expressed as

\begin{equation}
\Gamma _{3(T)}^{(1)\mu }=\frac{e^3}{8\pi \Theta }\int_0^1dx\int_0^1dy\left[ 
\frac{A^\mu }{\left( Q_0^2-y{\bf Q}^2\right) ^{3/2}}+\frac{B^\mu }{\left(
Q_0^2-y{\bf Q}^2\right) ^{1/2}}\right] \;,  \label{gama3T}
\end{equation}
with the numerators given by $A^0=\epsilon _{ij}p^iq^jQ^0\left(
P^0-2Q^0\right) $, $A^l=\epsilon _{ij}p^iq^jQ^0\left( P^l-2yQ^l\right) $, $%
B^0=\epsilon _{ij}p^iq^j\left( -2\right) $ and $B^l=\epsilon _{ij}\left(
4V^i+2Q^0q^i\right) g^{jl}$ where $q^\mu =p^\mu -p^{\prime \mu }$, $P^\mu
=p^\mu +p^{\prime \mu }$, $V^i=q^0p^i-p^0q^i$ and $Q^\mu (x)=p^\mu -q^\mu
\left( 1-x\right) $. The $y$ integration in (\ref{gama3T}) is easy but the remaining $
x $ one is very complicated. For sake of simplicity, we restrict ourselves
to the situation where $p^0=p^{\prime \,0}$ which the relevant one for the
calculation of the 1-loop CM two body scattering. In this case, $Q_0^2=m^2+
{\bf p}^2$ and expanding (\ref{gama3S}) and (\ref{gama3T}) for $|{\bf p}|/m$
small one obtains the total trilinear vertex correction, up to order ${\bf p}
^2/m^2$, as 
\begin{equation}
\left. \Gamma _3^{(1)0}(p,p-q)\right| _{q^0=0}\simeq \frac{-e^3}{4\pi \Theta 
}\left[ \frac{\epsilon _{ij}p^iq^j}m\right]  \label{gama30}
\end{equation}
and 
\begin{equation}
\left. \Gamma _3^{(1)l}(p,p-q)\right| _{q^0=0}\simeq \frac{-e^3}{4\pi \Theta 
}\left[ \epsilon _{ij}q^ig^{jl}\right] \left( 2+\frac 1{12}(5+\cos \theta )%
\frac{{\bf p}^2}{m^2}\right) +\frac{e^3}{4\pi \Theta }\left[ \frac{\epsilon
_{ij}p^iq^j}m\right] \frac{(p+p^{\prime })^l}{4m}\;,  \label{gama3l}
\end{equation}
where $\theta $ is the angle between the vectors ${\bf p}$ and ${\bf p}%
^{\prime }$. In the above equations, and from now on, the symbol $\simeq $
denotes that the expression which follows holds up to the order ${\bf p}%
^2/m^2$.

The one loop correction to the seagull vertex is of fourth order in the
charge $e$, and so it does not enter in the particle-particle scattering at
one loop level. The correction to the self-interaction vertex, which
actually represents the 2-particle scattering in one loop order, will be
calculated in the next section.

\section{Particle-Particle Scattering}

In the CM\ frame, with external particles on the mass shell, one has ${\bf p}%
_1=-{\bf p}_2={\bf p}$ , ${\bf p}_1^{\prime }=-{\bf p}_2^{\prime }={\bf p}%
^{\prime }$ and $p_1^0=p_2^0=p_1^{\prime \,0}=p_2^{\prime \,0}=w_p=\sqrt{m^2+%
{\bf p}^2\text{ }}$ . The tree level particle-particle amplitude, presented
in Fig. 3, is given by

\begin{equation}
A^{(0)}=-\lambda -i\frac{8e^2}\Theta \sqrt{m^2+{\bf p}^2}\cot \theta \simeq
-\lambda -i\frac{8e^2}\Theta m\left( 1+\frac{{\bf p}^2}{2m^2}\right) \cot
\theta \;,  \label{Atree}
\end{equation}
where $\theta $ is the scattering angle and $m$ is the renormalized mass of
the bosonic particle. One sees that, by definiteness, we take the amplitude
as being $(-i)$ times the 1PI four point function. This choice is only to
facilitate the comparison with the nonrelativistic case discussed in ref.[$%
\citen{bl}$].

We shall calculate the 1-loop order scattering amplitude, for low external
momenta and up to order ${\bf p}^2/m^2$. The insertion of the vacuum
polarization into the tree level gives, using (\ref{vacpolLQ}) with $q^0=0$,
the contribution

\begin{equation}
A^{(p)}\simeq -\frac{2e^4}{\pi \Theta ^2}m\left\{ \frac 16+\frac 7{30}\frac{%
{\bf p}^2}{m^2}\right\} \;.  \label{ApT}
\end{equation}
Similarly, the trilinear vertex correction $\Gamma_{3}^{(1)\mu }$, given by (%
\ref{gama30}) and (\ref{gama3l}), inserted into the tree level leads to the following $e^4$-contribution

\begin{equation}
A^{(v)}\simeq -\frac{2e^4}{\pi \Theta ^2}m\left\{ 2+\frac 53\frac{{\bf p}^2}{%
m^2}\right\} \;.  \label{AvT}
\end{equation}

Diagrams, like those shown in Fig. 4, that admixes particle self-interaction
and gauge field exchange, do not contribute. The first vanishes by charge
conjugation, the second is null due to the antisymmetric form of the gauge
field propagator whereas the chalice diagram is proportional to $\sin \theta 
$ and, thus, is eliminated by its final particles exchanged partner. The
most important, genuine, one loop particle-particle scattering comes from
the diagrams shown in Fig. 5 where it is also presented the routing of
external momenta used in the calculations.

The group $(a)$ is the finite self-interaction scattering, which can be
exactly calculated [$\citen{boz}$-$\citen{go}$] and is given by

\begin{eqnarray}
A^{(a)} &=&\frac{\lambda ^2}{32\pi m}\left\{ \frac 1{\sqrt{1+{\bf p}^2/m^2}}%
\left[ \ln \left( \frac{\sqrt{1+{\bf p}^2/m^2}+1}{\sqrt{1+{\bf p}^2/m^2}-1}%
\right) +i\pi \right] \right.  \nonumber   \\
&&\ \ \left. +\left[ \frac 2{\sqrt{(1-\cos \theta ){\bf p}^2/2m^2}}\arctan
\left( \sqrt{(1-\cos \theta ){\bf p}^2/2m^2}\right) +(\theta \rightarrow \pi
-\theta )\;\right] \right\}  \nonumber \\
\ &\simeq &\frac{\lambda ^2}{32\pi m}\left\{ \left( 1-\frac{{\bf p}^2}{2m^2}%
\right) \left[ \ln \left( \frac{4m^2}{{\bf p}^2}\right) +i\pi \right] +4-%
\frac{{\bf p}^2}{6m^2}\right\} .  \label{Aa}
\end{eqnarray}
The $|{\bf p}|/m$ expansion of the more involving CS scattering, the $(b)$
and $(c)$ groups of diagrams of Fig. 5, will be calculated employing the
following cutoff procedure [$\citen{go}$].

First of all, we integrate over $k^0$ (the frequency part of the loop
momentum $k$) without making any restriction in order to guarantee locality
in time. This integration is greatly facilitated in the gauge we are working
since the gauge field propagator does not depend on $k^0$. The remaining
integration over the Euclidean ${\bf k}$ plane is then separated into two
parcels through the introduction of an intermediate cutoff $\Lambda _I$ in
the $|{\bf k}|$ integration satisfying

\begin{equation}
(i)\;|{\bf p}|\ll \Lambda _I\ll m\hspace{0.5cm}\text{and}\hspace{0.5cm}%
(ii)\;\left( \frac{|{\bf p}|}{\Lambda _I}\right) ^2\approx \left( \frac{%
\Lambda _I}m\right) ^2\approx \frac{|{\bf p}|}m\;.  \label{cutoff}
\end{equation}
Condition $(i)$ presupposes that nonrelativistic spatial momenta are indeed
much smaller than the particle mass while condition $(ii)$ establishes $\eta
\approx |{\bf p}|/m$ as the small expansion parameter. The auxiliary cutoff $%
\Lambda _I$ splits the space of the intermediate states into two parts, the
low (L) energy sector $\left( |{\bf k}|<\Lambda _I\right) $ and the high (H)
energy one with $|{\bf k}|$ $>\Lambda _I$. In the L sector all the spatial
momenta involved are small ($|{\bf p}|/m$ $,$ $|{\bf k}|/m\ll 1$) and so one
can perform a $1/m$ expansion of the integrand while, in the H- region, $|%
{\bf k}|\gg |{\bf p}|$ and the integrand can be expanded in a Taylor series
around $|{\bf p}|=0$ and then, in both cases, integrated term by term (a
regularization scheme has to be used if the graph is ultraviolet divergent).
This procedure permits analytical calculations in every order in $\eta $,
produces $\Lambda _I$- dependent results and further expansions in $\Lambda
_I/m$ may be necessary to get the $|{\bf p}|/m$ expansion of the L and H
contributions to the amplitude, up to the desired order. Certainly, for sake
of consistence, the $\Lambda _I$- dependent parcels of the L and the H
contributions of each diagram cancel identically. This process has been
explicitly verified to produce the correct $|{\bf p}|/m$ expansion for the
self-interaction scattering $(a)$ [$\citen{go}$].

Consider the ``right'' box diagram corresponding to the direct exchange of
two virtual gauge particles, the first parcel of Fig. 5$(b)$. Following the
Feynman rules, one has

\begin{eqnarray}
A^{(b)dir} &=&-ie^4\int \frac{d^3k}{(2\pi )^3}\left\{ (p_1+k)^\mu \,D_{\mu
\sigma }(k-p_1)\,(2p_2+p_1-k)^\sigma \,\Delta (k)\right.  \nonumber \\
&&\ \ \ \left. \Delta (p_1+p_2-k)\,(-k+p_1+p_2+p_2^{\prime })^\rho \,D_{\rho
\nu }(k-p_1^{\prime })\,(k+p_1^{\prime })^\nu \right\} \,+\,\left[
p_1^{\prime }\leftrightarrow p_2^{\prime }\right]  \nonumber \\
\ &=&-\frac{4e^4}{\pi ^2\Theta ^2}\int d^2{\bf k\,}\left( \frac{w_p^2}{w_k}%
\right) \,\frac 1{{\bf p}^2-{\bf k}^2+i\varepsilon }\,\left[ \frac{({\bf %
k\times p})\,({\bf k\times p}^{\prime })}{({\bf k-p})^2({\bf k-p}^{\prime
})^2}\right] \,+\,\left[ {\bf p}^{\prime }\leftrightarrow -{\bf p}^{\prime
}\right] \;,  \label{abd}
\end{eqnarray}
where the $k^0$ integration was done as a contour integral. The angular
integration in the last line above can be cast in the form $\frac 12\left[
\cos \theta \,{\rm I}_0-{\rm I}_2\right] $ where

\begin{equation}
{\rm I}_n=\int_0^{2\pi }d\varphi \,\frac{\cos \,(n\varphi )}{\left[ 2\cos
(\varphi -\theta /2)-\beta \right] \left[ 2\cos (\varphi +\theta /2)-\beta
\right] }  \label{In}
\end{equation}
and $\beta =({\bf k}^2+{\bf p}^2)/(|{\bf k}||{\bf p}|)$ . This integral can
be done using the residue theorem and one finds

\begin{eqnarray}
A^{(b)dir} &=&-\frac{e^4}{\pi \Theta ^2}\int d({\bf k}^2)\ \left( \frac{w_p^2%
}{w_k}\right) \frac 1{{\bf p}^2-{\bf k}^2+i\varepsilon }  \nonumber \\
&&\ \hspace{2.5cm}\left[ \frac{|\,{\bf k}^2-{\bf p}^2\,|\,(\,{\bf k}^2-{\bf p%
}^2\,)}{({\bf k}^2)^2+({\bf p}^2)^2-2\,{\bf k}^2\,{\bf p}^2\,\cos \theta }%
-1\right] \,+\,\left( \theta \leftrightarrow \pi -\theta \right) \;.
\label{abd2}
\end{eqnarray}
The remaining ${\bf k}^2$ integration is then divided into two pieces, from $%
0$ to $\Lambda_{I}^{2}$ (L region) and from $\Lambda_I ^2$ to $\Lambda
_0^2\rightarrow \infty $ (H sector). In the L part, using

\begin{equation}
\frac{w_p^2}{w_k}=m\left( 1+\frac{{\bf p}^2}{m^2}\right) \left[ 1-\frac{{\bf %
k}^2}{2m^2}+\frac{3({\bf k}^2)^2}{8m^4}+...\right]  \label{wpwk}
\end{equation}
and keeping terms up to order $\eta ^2$, one obtains

\begin{eqnarray}
A_L^{(b)dir} &\simeq &-\frac{2e^4}{\pi \Theta ^2}m\left\{ \left( 1+\frac{%
{\bf p}^2}{2m^2}\right) \left[ \ln (2|\sin \theta |)+i\pi \right] -\frac{%
{\bf p}^2}{m^2}\right.  \nonumber \\
&&\ \ \left. \hspace{1.5cm}-\frac 12\cos \theta \ln \left( \frac{1-\cos
\theta }{1+\cos \theta }\right) \frac{{\bf p}^2}{m^2}-(1-2\cos ^2\theta )%
\frac{{\bf p}^4}{\Lambda_I ^4}\right\} \;.  \label{AbdirL}
\end{eqnarray}
In the H region, the integrand is replaced by its Taylor expansion around $%
{\bf p}^2=0$ which is given by 
\[
\left[ -\frac{2m^2\cos \theta }{({\bf k}^2)^2\sqrt{{\bf k}^2+m^2}}\right] 
{\bf p}^2+\left[ \frac{2m^2(1-2\cos ^2\theta )}{({\bf k}^2)^3\sqrt{{\bf k}%
^2+m^2}}-\frac{2\cos \theta \sqrt{{\bf k}^2+m^2}}{({\bf k}^2)^3}\right] {\bf %
p}^4+{\cal O}({\bf p}^6)\;. 
\]
Performing the ${\bf k}^2$ integrations one obtains, up to order $\eta ^2$,

\begin{equation}
A_H^{(b)dir}\simeq -\frac{2e^4}{\pi \Theta ^2}m\left\{ (1-2\cos ^2\theta )%
\frac{{\bf p}^4}{\Lambda_I ^4}\right\} \;.  \label{AbdirH}
\end{equation}
Adding (\ref{AbdirL}) and (\ref{AbdirH}), we get

\begin{eqnarray}
A^{(b)dir} &\simeq &-\frac{2e^4}{\pi \Theta ^2}m\left\{ \left( 1+\frac{{\bf p%
}^2}{2m^2}\right) \left[ \ln (2|\sin \theta |)+i\pi \right] -\frac{{\bf p}^2%
}{m^2}\right.  \nonumber   \\
&&\ \ \left. \hspace{1.5cm}-\frac 12\cos \theta \ln \left( \frac{1-\cos
\theta }{1+\cos \theta }\right) \frac{{\bf p}^2}{m^2}\right\} \;.
\label{Abdir}
\end{eqnarray}

The $k^0$ integration of the twisted box diagram, the second in Fig. 5$(b)$,
gives

\begin{eqnarray}
A^{(b)twist} &=&-\frac{e^4}{2\pi ^2\Theta ^2}\int d^2{\bf k\,}\left( \frac{%
w_kw_{k-s}-w_p^2}{w_kw_{k-s}(w_k+w_{k-s})}\right)  \nonumber \\
&&\ \ \ \ \hspace{2.5cm}\left[ \frac{({\bf k\times s})^2-({\bf p\times p}%
^{\prime })^2}{({\bf k-p})^2({\bf k-p}^{\prime })^2}\right] \,+\,\left[ {\bf %
p}^{\prime }\leftrightarrow -{\bf p}^{\prime }\right]  \label{abt}
\end{eqnarray}
where ${\bf s=p}+{\bf p}^{\prime }$ and we recall that $w_{k-s}=\sqrt{({\bf k%
}-{\bf s})^2+m^2}$ . This integration has a rather non trivial angular part
but the use of the approximation procedure before performing it allows
analytical calculations and one ends up, after adding its final particles
exchanged partner, with

\begin{equation}
A_L^{(b)twist}\simeq 0\;\;\text{and}\;\;A^{(b)twist}\simeq
A_H^{(b)twist}\simeq -\frac{2e^4}{\pi \Theta ^2}m\left\{ \frac{{\bf p}^2}{%
2m^2}\right\} \;.  \label{Abtwist}
\end{equation}
Thus, the total box amplitude, $A^{(b)}=A^{(b)dir}+A^{(b)twist}$, is finite
and, up to order ${\bf p}^2/m^2$, is given by

\begin{eqnarray}
\ A^{(b)} &\simeq &-\frac{2e^4}{\pi \Theta ^2}m\left\{ \left( 1+\frac{{\bf p}%
^2}{2m^2}\right) \left[ \ln (2|\sin \theta |)+i\pi \right] -\frac{{\bf p}^2}{%
2m^2}\right.  \nonumber \\
&&\ \ \ \left. \hspace{1.5cm}-\frac 12\cos \theta \ln \left( \frac{1-\cos
\theta }{1+\cos \theta }\right) \frac{{\bf p}^2}{m^2}\right\} \;.
\label{Abox}
\end{eqnarray}
Notice that, as a by product, this cutoff procedure gives the origin
(whether from the L or H sectors) of each contribution. One sees, for
example, that the contribution of the twisted box diagram comes entirely
from H energy which is naturally expected from a graph that involves
backward propagation in time.

The third group, the seagull scattering, has to be treated more carefully
since it carries the divergence of the four point function. One can
immediately see that each of the $k^0$ integrations of the the gauge bubble
and the two triangle diagrams would diverge if made separately. However,
taking all the diagrams of group $(c)$ together, the divergences of the $k^0$
integrations cancel out identically and the angular integrations, which
again are linear combinations of ${\rm I}_n\,$, lead to

\begin{eqnarray}
A^{(c)} &=&-\frac{e^4}{2\pi \Theta ^2}\left\{ \int d({\bf k}^2)\,\left( 
\frac{w_p^2+w_k^2}{w_k}\right) \,\frac{{\rm sgn}({\bf k}^2{\bf -p}^2)\,(\,%
{\bf k}^2-\cos \theta \,{\bf p}^2\,)}{({\bf k}^2)^2+({\bf p}^2)^2-2\,{\bf k}%
^2\,{\bf p}^2\,\cos \theta }\right.  \nonumber \\
&&\ \ \ \ \ \left. \hspace{1.0cm}-\int d({\bf k}^2){\bf \,}\frac 1{w_k}%
\,\left[ \frac{|\,{\bf k}^2-{\bf p}^2\,|\,(\,{\bf k}^2-{\bf p}^2\,)}{({\bf k}%
^2)^2+({\bf p}^2)^2-2\,{\bf k}^2\,{\bf p}^2\,\cos \theta }-1\right] \right\}
\,+\,\left( \theta \leftrightarrow \pi -\theta \right).  \label{acf}
\end{eqnarray}
Repeating the procedure exemplified with the box diagram one finds

\begin{eqnarray}
A_L^{(c)} &\simeq &-\frac{2e^4}{\pi \Theta ^2}m\left\{ \left( 1+\frac{{\bf p}%
^2}{2m^2}\right) \left[ \ln \left( \frac{\Lambda_I ^2}{{\bf p}^2}\right) -\ln
(2|\sin \theta |)\right] +\frac{{\bf p}^2}{m^2}\right.  \nonumber \\
&&\ \ \ \left. \hspace{1.5cm}+\frac 12\cos \theta \ln \left( \frac{1-\cos
\theta }{1+\cos \theta }\right) \frac{{\bf p}^2}{m^2}+\frac 12(1-2\cos
^2\theta )\frac{{\bf p}^4}{\Lambda_I ^4}+\frac{\Lambda_I ^4}{16m^4}\right\} \;,
\label{AcL}
\end{eqnarray}

\begin{eqnarray}
A_H^{(c)} &\simeq &-\frac{2e^4}{\pi \Theta ^2}m\left\{ -\left( 1+\frac{{\bf p%
}^2}{2m^2}\right) \ln \left( \frac{\Lambda_I ^2}{4m^2}\right) -1\right. 
\nonumber \\
&&\ \ \ \left. \hspace{1.5cm}-\frac 12(1-2\cos ^2\theta )\frac{{\bf p}^4}{%
\Lambda_I ^4}-\frac{\Lambda_I ^4}{16m^4}+\left[ \frac{\Lambda _0}m\right]
\right\}  \label{AcH}
\end{eqnarray}

\noindent and, thus, the total seagull contribution to the amplitude is

\begin{eqnarray}
A^{(c)} &\simeq &-\frac{2e^4}{\pi \Theta ^2}m\left\{ \left( 1+\frac{{\bf p}^2%
}{2m^2}\right) \left[ \ln \left( \frac{4m^2}{{\bf p}^2}\right) -\ln (2|\sin
\theta |)\right] \right.  \nonumber \\
&&\ \ \ \ \ \ \left. \hspace{1.5cm}-1+\frac 12\cos \theta \ln \left( \frac{%
1-\cos \theta }{1+\cos \theta }\right) \frac{{\bf p}^2}{m^2}+\frac{{\bf p}^2%
}{m^2}\right\} +\frac{2e^4}{\pi \Theta ^2}\Lambda _0\;.  \label{Ac}
\end{eqnarray}
The constant divergent term above can be suppressed by a counterterm of the
form $-\frac{2e^4}{\pi \Theta ^2}\Lambda _0(\phi ^{*}\phi )^2$ introduced in
the Lagrangian density. We can also imagine that the bare self-coupling $%
\lambda $ carries a divergent part that just cancel the divergence of the
four point function. In any case, we take the finite part of (\ref{Ac}) as
the 1-loop renormalized $(c)$ contribution. This would be the result if we
had used dimensional renormalization.

The pure CS exchange scattering, the sum $A^{(b)}+A^{(c)}$, is given by 
\begin{equation}
A^{(CS)}\simeq -\frac{2e^4}{\pi \Theta ^2}m\left\{ \left( 1+\frac{{\bf p}^2}{%
2m^2}\right) \left[ \ln \left( \frac{4m^2}{{\bf p}^2}\right) +i\pi \right]
-1+\frac{{\bf p}^2}{2m^2}\right\} \;,  \label{ACS}
\end{equation}
and it is noticeable that the cancellation of the $\theta $ dependent terms
of the box and the seagull amplitudes happens in both dominant and
subleading orders.

The total renormalized 1-loop particle-particle scattering amplitude, $%
A^{(a)}+A^{(CS)}+A^{(p)}+A^{(v)}$, is independent of the scattering angle $%
\theta $ and, up to order ${\bf p}^2/m^2$, is given by

\begin{eqnarray}
\ A^{(1)} &\simeq &m\left( \frac{\lambda ^2}{32\pi m^2}-\frac{2e^4}{\pi
\Theta ^2}\right) \left[ \ln \left( \frac{4m^2}{{\bf p}^2}\right) +i\pi
\right]  \nonumber \\
&&\ \ -m\left( \frac{\lambda ^2}{32\pi m^2}+\frac{2e^4}{\pi \Theta ^2}%
\right) \frac{{\bf p}^2}{2m^2}\left[ \ln \left( \frac{4m^2}{{\bf p}^2}%
\right) +i\pi \right]  \nonumber \\
&&\ \ +m\left( \frac{\lambda ^2}{8\pi m^2}-\frac{7e^4}{3\pi \Theta ^2}%
\right) -m\left( \frac{\lambda ^2}{192\pi m^2}+\frac{24e^4}{5\pi \Theta ^2}%
\right) \frac{{\bf p}^2}{m^2}\;.  \label{Atotal}
\end{eqnarray}
The leading term of the above expansion, which coincides with the result of
Ref. [$\citen{boz}$], vanishes if the self-interaction parameter is fixed at
one of the critical values $\lambda _c^{\pm }=\pm 8me^2/|\Theta |\,$ but the
subdominant terms do not. The implications of this fact will be discussed
next.

\section{Relativistic Corrections to AB Scattering}

Prior to any comparison with the nonrelativistic case, the normalization of
states has to be properly adjusted. In the relativistic case one takes $%
\left\langle {\bf p}^{\prime }|{\bf p}\right\rangle =2w_p\delta ({\bf p}%
^{\prime }-{\bf p})$ while the usual normalization in a NR theory does not
have the $2w_p$ factor and thus the CM\ amplitudes, calculated in the last
section, must be multiplied by 
\begin{equation}
\left( \frac 1{\sqrt{2w_p}}\right) ^4=\frac 1{4m^2}\left[ 1-\frac{{\bf p}^2}{%
m^2}+...\right] \;.  \label{norfac}
\end{equation}
The tree level and the 1-loop amplitudes, equations (\ref{Atree}) and ( \ref
{Atotal}), are then rewritten, up to order ${\bf p}^2/m^2$, as

\begin{equation}
{\cal A}^{(0)}=-\frac \lambda {4m^2}-i\frac{2e^2}{m\Theta }\cot \theta
+\left[ \frac \lambda {4m^2}+i\frac{e^2}{m\Theta }\cot \theta \right] \frac{%
{\bf p}^2}{m^2}  \label{AtreeNR}
\end{equation}
and

\begin{eqnarray}
{\cal A}^{(1)} &\simeq &\frac m{8\pi }\left( \frac{\lambda ^2}{16m^4}-\frac{%
4e^4}{m^2\Theta ^2}\right) \left[ \ln \left( \frac{4m^2}{{\bf p}^2}\right)
+i\pi \right]  \nonumber \\
&&\ \ \ -\frac m{8\pi }\left( \frac{3\lambda ^2}{32m^4}-\frac{2e^4}{%
m^2\Theta ^2}\right) \frac{{\bf p}^2}{2m^2}\left[ \ln \left( \frac{4m^2}{%
{\bf p}^2}\right) +i\pi \right]  \nonumber \\
&&\ \ \ +\frac m{8\pi }\left( \frac{\lambda ^2}{4m^4}-\frac{14e^4}{%
3m^2\Theta ^2}\right) -\frac m{8\pi }\left( \frac{25\lambda ^2}{96m^4}+\frac{%
74e^4}{15m^2\Theta ^2}\right) \frac{{\bf p}^2}{m^2}\;,  \label{AtotalNR}
\end{eqnarray}
where calligraphic ${\cal A}$ means that the amplitude is written in the
nonrelativistic normalization.

Confronting the tree levels, of the relativistic and the NR scattering
amplitudes, one sees that the self-interaction parameters are related by $%
v=\lambda /4m^2\,$and the critical values for which the 1-loop NR and
leading relativistic scattering amplitudes vanish are also related by

\begin{equation}
v_c^{\pm }=\frac{\lambda _c^{\pm }}{4m^2}=\pm \frac{2e^2}{m|\Theta |}=\pm 
\frac{4\pi }m\,\alpha \,{\rm sgn\,}(\alpha )\;,  \label{critic}
\end{equation}
where the AB parameter is $\alpha =e^2/2\pi \Theta $. By choosing the value $%
v_c^{+}$, corresponding to a repulsive contact interaction, the tree
amplitude reduces, after multiplying by the appropriated kinematical factor,
to the Aharonov-Bohm amplitude for identical particles (\ref{AB}). The
leading order vanishes at $v_c^{+}$ whereas the subdominant terms that
survive, namely

\begin{eqnarray}
{\cal A}^{{\rm sub}} &=&\frac{4\pi }m\,\alpha \,\left[ {\rm sgn\,}(\alpha )+%
\frac i2\cot \theta \right] \frac{{\bf p}^2}{m^2}+\frac{17}3\frac \pi m%
\,\alpha ^2  \nonumber   \\
&&\ -\frac \pi m\,\alpha ^2\frac{{\bf p}^2}{m^2}\left[ \ln \left( \frac{4m^2%
}{{\bf p}^2}\right) +i\pi \right] -\frac{54}5\frac \pi m\,\alpha ^2\frac{%
{\bf p}^2}{m^2}\;,  \label{Asubled}
\end{eqnarray}
represent relativistic corrections to the Aharonov-Bohm scattering.

Part of the correction of the tree level $(\,\sim \alpha )$ is due to the
normalization of states and so has a pure kinematical origin, but not all of
it since the scattering amplitude corresponding to the exchange of one
virtual gauge particle depends on the CM energy as a consequence of the
minimal coupling. The other corrections come from the 1-loop $(e^4)$
contribution to the perturbative expansion and are indeed relativistic.
These kind of terms, proportional to $\alpha ^2$, do not exists in
nonrelativistic AB scattering (which exact result is function of $\sin
\alpha $ ) and then may be detected in experiments with fast particles. The
relativistic correction has non trivial contributions (the third parcel of (%
\ref{Asubled}), for example) that can only be incorporated in a
nonrelativistic framework through the addition of new nonrenormalizable
interactions to the Lagrangian (\ref{lagranNR}). This aspect will be
discussed elsewhere.
\newpage
\begin{flushleft}
{\bf Acknowledgements}
\end{flushleft}
This work was partially supported by Conselho Nacional de
Desenvolvimento Cient\'\i fico e Tecnol\'ogico (CNPq) e Funda\c c\~ao de
Amparo \`a Pesquisa do Estado de S\~ao Paulo (FAPESP).

\newpage

\begin{center}
Figure captions
\end{center}

\vspace{2.0cm}

Fig. 1 - Vacuum polarization correction in one loop order.

\vspace{1.0cm}

Fig. 2 - One loop correction to the trilinear vertex.

\vspace{1.0cm}

Fig. 3 - Tree level scattering.

\vspace{1.0cm}

Fig. 4 - Basic mixing interactions diagrams .

\vspace{1.0cm}

Fig. 5 - One loop order particle-particle scattering. In the momenta
assignment shown, $s=p_1+p_2$ , $q=p_1-p_1^{\prime }$ and $u=p_1^{\prime
}-p_2$ .

\end{document}